\documentclass[manuscript, screen]{acmart}

\usepackage{multirow}
\usepackage{graphicx}
\usepackage{xspace}
\usepackage{tcolorbox}
\usepackage{booktabs}
\usepackage{enumitem}
\usepackage{tikz}
\usepackage{xcolor}
\usepackage{pifont}
\usepackage{amsmath}
\usepackage[table]{xcolor} % For coloring table cells
\usepackage{tabularx}      % For creating tables of a set width
\usepackage{booktabs}
\usepackage{bbding}

\AtBeginDocument{%
  }

\setcopyright{acmlicensed}
\acmJournal{TOSEM}

\newcommand{\toolname}{LLM-as-a-Judge\xspace}

\renewcommand\footnotetextcopyrightpermission[1]{}
\begin{document}

\title[From Code to Courtroom: LLMs as the New Software Judges]{From Code to Courtroom: LLMs as the New Software Judges}

\author{Junda He}
\email{jundahe.2022@phdcs.smu.edu.sg}

\author{Jieke Shi}
\email{jiekeshi@smu.edu.sg}

\affiliation{%
  \institution{Singapore Management University}
  \country{Singapore}
}

\author{Terry Yue Zhuo}\authornote{Corresponding author}
\email{terry.zhuo@monash.edu}

\affiliation{%
  \institution{Monash University \& CSIRO's Data61}
  \country{Australia}
}

\author{Christoph Treude}
\email{ctreude@smu.edu.sg}

\affiliation{%
  \institution{Singapore Management University}
  \country{Singapore}
}

\author{Jiamou Sun}
\email{frank.sun@anu.edu.au}
\affiliation{%
\institution{CSIRO's Data61} 
\country{Australia}
}

\author{Zhenchang Xing}
\email{zhenchang.xing@anu.edu.au}

\affiliation{%
\institution{CSIRO's Data61 \& Australian National University} 
\country{Australia}
}

\author{Xiaoning Du}
\email{xiaoning.du@monash.edu}
\affiliation{%
  \institution{Monash University}
  \country{Australia}
}

\author{David Lo}
\email{davidlo@smu.edu.sg}
\affiliation{%
  \institution{Singapore Management University}
  \country{Singapore}
}

\renewcommand{\shortauthors}{He et al.}

\begin{abstract}
The rapid integration of Large Language Models (LLMs) into software engineering (SE) has revolutionized tasks from code generation to program repair, producing a massive volume of software artifacts. This surge in automated creation has exposed a critical bottleneck: the lack of scalable and reliable methods to evaluate the quality of these outputs. Human evaluation, while effective, is very costly and time-consuming. Traditional automated metrics like BLEU rely on high-quality references and struggle to capture nuanced aspects of software quality, such as readability and usefulness. In response, the LLM-as-a-Judge paradigm, which employs LLMs for automated evaluation, has emerged. This approach leverages the advanced reasoning and coding capabilities of LLMs themselves to perform automated evaluations, offering a compelling path toward achieving both the nuance of human insight and the scalability of automated systems. Nevertheless, LLM-as-a-Judge research in the SE community is still in its early stages, with many breakthroughs needed. 

This forward-looking SE 2030 paper aims to steer the research community toward advancing LLM-as-a-Judge for evaluating LLM-generated software artifacts, while also sharing potential research paths to achieve this goal. We provide a literature review of existing SE studies on LLM-as-a-Judge and envision these frameworks as reliable, robust, and scalable human surrogates capable of evaluating software artifacts with consistent, multi-faceted assessments by 2030 and beyond. To validate this vision, we analyze the limitations of current studies, identify key research gaps, and outline a detailed roadmap to guide future developments of LLM-as-a-Judge in software engineering. While not intended to be a definitive guide, our work aims to foster further research and adoption of LLM-as-a-Judge frameworks within the SE community, ultimately improving the effectiveness and scalability of software artifact evaluation methods.

  \end{abstract}

\settopmatter{printacmref=false}
\renewcommand\footnotetextcopyrightpermission[1]{}
\pagestyle{plain}



%%
%% Keywords. The author(s) should pick words that accurately describe
%% the work being presented. Separate the keywords with commas.
%% A "teaser" image appears between the author and affiliation
%% information and the body of the document, and typically spans the
%% page.

\received{20 February 2025}
\received[revised]{}
\received[accepted]{}

%%
%% This command processes the author and affiliation and title
%% information and builds the first part of the formatted document.
\maketitle

\section{Introduction}

Recent advances in Large Language Models (LLMs)~\cite{OpenAI_GPT4, deepseek2405deepseek, anthropic2023claude} are transforming the field of Software Engineering (SE)~\cite{he2025llm,10.1145/3712005,fan2023large,wang2024software, hu2025assessing}. These models have demonstrated exceptional performance on longstanding challenges, from code generation~\cite{zhuo2024bigcodebench, liu2024exploring} and code summarization~\cite{ahmed2024automatic, sun2024source} to program translation~\cite{yuan2024transagent, pan2023understanding} and repair~\cite{10.1109/ICSE48619.2023.00125,jin2023inferfix}.
Consequently, LLMs have powered a series of new applications that streamline the software development process~\cite{githubGitHubCopilot, cursor}.

However, the proliferation of LLMs in software development has led to a surge of generated software artifacts. Evaluating this massive output is a non-trivial task, as the inherent complexity and open-endedness of these artifacts present a profound assessment challenge. This raises a new challenging question for software engineering research:

\begin{tcolorbox}[width=\textwidth, box align=center, left=1.2cm, top=0pt, bottom=0pt, right=0.5cm, colback=white, frame empty]
{\it How can we comprehensively and scalably evaluate the quality of LLM-generated software artifacts?}
\end{tcolorbox}

While human evaluation by experienced developers remains the ideal approach for assessing the quality and effectiveness of these LLM-generated software artifacts, rigorous human evaluation faces significant practical hurdles.
A survey of SE researchers revealed that 84\% agree that human evaluation is problematic, primarily due to time constraints, the need for specialized practical knowledge, and the high cost involved~\cite{buse2011benefits}. Additionally, human evaluators may experience fatigue and reduced concentration, which further undermines the quality of the evaluation and delays the evaluation process~\cite{kumar2024llms}.

To overcome these challenges, researchers often turn to automated metrics~\cite{hu2022correlating,papineni2002bleu, lin2004rouge}, which offer scalability at a lower cost. Yet, traditional automated metrics have their own drawbacks. For example, Pass@$k$ is a widely used metric for measuring code correctness in code generation, which involves executing the first $k$ generated code snippets against unit tests. Although useful, Pass@$k$ demands substantial human effort to design comprehensive test suites and configure execution environments. For other tasks like code summarization and translation, researchers often use text-similarity metrics (e.g., BLEU~\cite{papineni2002bleu}, CodeBLEU~\cite{ren2020codebleu}) and embedding-based metrics (e.g., BERTScore~\cite{zhang2019bertscore}, CodeBERTScore~\cite{zhou2023codebertscore}). However, the effectiveness of these metrics also relies heavily on high-quality references that are typically annotated by human experts.

More broadly, traditional automated metrics are ill-equipped to handle the inherent complexity of LLM-generated content. They struggle to assess the nuanced aspects, such as naturalness, usefulness, and adherence to best practices. 
This fundamental limitation causes their results to frequently misalign with human judgment, a finding confirmed by several studies~\cite{hu2022correlating, kumar2024llms, roy2021reassessing, evtikhiev2023out}.
Consequently, achieving comprehensive evaluation while ensuring scalable automation remains a persistent challenge in assessing the quality of LLM-generated software artifacts.

Recently, the remarkable success of LLMs has inspired the emergence of the LLM-as-a-Judge (LLM-J) paradigm~\cite{zheng2024judging}. This paradigm extends LLMs' role from content generation to content evaluation, positioning them as scalable and cost-effective surrogates for human experts. Several key attributes make LLMs particularly well-suited for this evaluation role. First, numerous studies have shown that LLMs exhibit both impressive coding abilities~\cite{liu2024deepseek, lozhkov2024starcoder, li2023starcoder} and human-like reasoning skills~\cite{patil2025advancing, ivanova2025evaluate,guo2025deepseek}. 
Additionally, LLMs are often trained through reinforcement learning from human feedback (RLHF)~\cite{bai2022training}, which helps in closely aligning LLM's judgments with human expert judgments~\cite{chen2021evaluating}. Moreover, unlike human evaluators, LLMs do not experience fatigue or reduced efficiency from prolonged work, allowing them to maintain consistent performance over extended periods. Together, these attributes collectively make LLMs compelling candidates for automated evaluation.

The SE community is increasingly exploring LLM-as-a-Judge to overcome the limitations of both costly human evaluation and traditional automated metrics. A growing body of research work demonstrates the promise of this approach~\cite{wang2025can,wu2024can,ahmed2024can,zhuo2023ice}. Nevertheless, the research area remains in its early stages, and significant breakthroughs are required to realize the full potential of this paradigm by 2030 and beyond. Therefore, this paper aims to present a forward-looking vision and a concrete research roadmap that inspires further progress.

In the following sections, this paper first provides a formal definition of the \toolname concept (\autoref{sec:background}). We then review key historical milestones and recent advances in \toolname for software engineering (\autoref{sec:review}). Looking beyond 2030, we envision \toolname systems as reliable, robust, and scalable human surrogates capable of evaluating a wide range of software artifacts while providing consistent, multi-faceted assessments.
To validate this vision, we identify key limitations and research gaps, outlining a roadmap with specific research directions and potential solutions for the SE research community to pursue (\autoref{sec:roadmap}). In summary, our study makes the following key contributions:
\begin{itemize}[leftmargin=*]
    \item A systematic review of 42 primary studies on the use of \toolname in software engineering, providing a comprehensive summary of its applications.
    \item An analysis of the limitations of current research, identifying key challenges and research gaps in the field.
    \item A forward-looking vision for \toolname in SE toward 2030 and beyond, supported by a detailed research roadmap to guide future exploration.
\end{itemize}
\section{Definition of LLM-as-a-Judge}
\label{sec:background}

As a recent emerging topic, various definitions of \toolname are currently proposed and not yet unified. 
This paper was inspired by the definition proposed by Li et al.~\cite{li2024llms}. 
Informally speaking, \toolname utilizes LLMs as evaluative tools to assess software artifact based on predefined evaluation criteria.
Formally speaking, \toolname is defined as follows:

\begin{equation}
 E(\mathcal{T}, \mathcal{C}, \mathcal{X}, \mathcal{R}) \rightarrow (\mathcal{Y}, \mathcal{E}, \mathcal{F})
\end{equation}

The function \( E \) is the core evaluation mechanism powered by LLMs. It takes the following key \textbf{inputs}:

\begin{itemize}[leftmargin=*]
    \item \( \mathcal{T} \) – Evaluation Type. It mainly covers three types: \textit{point-wise}, \textit{pairwise}, and \textit{list-wise}.
    \begin{itemize}
        \item \textbf{Point-wise Evaluation:} Function \( E \) evaluates a single candidate content independently and often provides a score or categorical classification.
        \item \textbf{Pair-wise Evaluation:} Function \( E \) compares two candidates and determines their relative quality.
        \item \textbf{List-wise Evaluation:} Function \( E \) ranks multiple (more than 2) candidates.
    \end{itemize}
    \item \( \mathcal{C} \) – Evaluation Criteria. It describes the assessment aspects, e.g., correctness, helpfulness, readability, etc.
    \item \( \mathcal{X} \) – Evaluation Item. It is the content to be judged. Note that \( \mathcal{X} \) can be a single candidate or a set of candidates \( (x_1, x_2, ..., x_n) \), where \( x_i \) is the \( i^{th} \) candidate to be evaluated.
    \item \( \mathcal{R} \) – Reference (optional). For instance, for a code generation task, \( \mathcal{R} \) can be a code implementation that successfully fulfills the requirements of a coding task. Note that \( \mathcal{R} \) is not mandatory, many \toolname systems operate in a reference-free manner.

\end{itemize}

Given these inputs, \( E \) can produce three primary \textbf{outputs}:

\begin{itemize}[leftmargin=*]
    \item \( \mathcal{Y} \) – Evaluation Result. \( \mathcal{Y} \) can take one of the following forms:
    \begin{itemize}[leftmargin=*]
        \item \textbf{Graded Evaluation:} Each candidate is assigned a score, either numerical (discrete or continuous) or categorical.      
        \item \textbf{Rank Ordering:} Candidates  \( (x_1, x_2, ..., x_n) \) are ranked based on their assessed quality.
        \item \textbf{Best-Choice Selection:} Function \( E \) selects the most suitable candidate ($x_i$) from the given set \( (x_1, x_2, ..., x_n) \).
    \end{itemize}

    \item \( \mathcal{E} \) – Evaluation Explanation. Optionally provides reasoning and justification for the evaluation.  
    \item \( \mathcal{F} \) – Feedback. Optionally offers constructive suggestions for improving the evaluated item \( \mathcal{X} \).  
\end{itemize}

In general, we envision \toolname{} as a genuine surrogate for human evaluation. It should support a wide spectrum of nuanced assessments (e.g., usefulness, adequacy, and conciseness) and ideally provide justifications or constructive feedback. 
\section{Review Methodology}
\label{sec:review-methodology}

This section presents the methodology of our systematic literature review. We first describe the survey scope along with the inclusion and exclusion criteria, then introduce the data sources, search strategy, and snowballing procedure used to achieve comprehensive coverage. Finally, we outline the quality assessment criteria applied to the selected studies.

\subsection{Survey Scope}

The scope of this review is defined by two primary dimensions: the research must focus on software engineering and must employ the LLM-as-a-Judge paradigm.

\vspace{1mm}
\noindent\textbf{Inclusion/Exclusion Criteria}. To ensure the selection of relevant and high-quality studies, we established the following inclusion and exclusion criteria:

\begin{itemize}[leftmargin=1.5em]
\item[{\textcolor[RGB]{0,128,0}{\Checkmark}}] \emph{The paper must be written in English.}
\item[{\textcolor[RGB]{0,128,0}{\Checkmark}}] \emph{The full text of the paper must be accessible, either through open-access channels or the researchers' institutional subscriptions.}
\item[{\textcolor[RGB]{0,128,0}{\Checkmark}}] \emph{The paper must utilize the LLM-as-a-Judge paradigm to one or more software engineering tasks.}
\item[{\textcolor[RGB]{209,26,66}{\XSolidBrush}}] \emph{The paper has less than 5 pages.}
\item[{\textcolor[RGB]{209,26,66}{\XSolidBrush}}] \emph{The paper is a duplicate or a minor extension of another study by the same authors.}
\item[{\textcolor[RGB]{209,26,66}{\XSolidBrush}}] \emph{Books, keynote records, panel summaries, technical reports, theses, tool demo papers, editorials.}
\item[{\textcolor[RGB]{209,26,66}{\XSolidBrush}}] \emph{The paper is a literature review or survey.}
\item[{\textcolor[RGB]{209,26,66}{\XSolidBrush}}] \emph{The paper does not involve software engineering-related tasks.}
\item[{\textcolor[RGB]{209,26,66}{\XSolidBrush}}] \emph{The paper lacks experimental results and mentions the LLM-as-a-Judge approach only in future work or discussions.}
\end{itemize}

\subsection{Paper Source}

For our literature search, we selected the DBLP computer science bibliography\footnote{\url{https://dblp.org/}} as the primary data source. This choice follows established practices in systematic literature reviews (SLR) for software engineering~\cite{he2025llm,liu2024large}. Studies~\cite{zhang2020machine} have shown that DBLP's corpus broadly subsumes other major academic databases, such as IEEE Xplore\footnote{\url{https://ieeexplore.ieee.org/}} and the ACM Digital Library.\footnote{\url{https://dl.acm.org/}} A key advantage of DBLP is its indexing of arXiv,\footnote{\url{https://arxiv.org/}} which allows the inclusion of recent preprints that may face delays before appearing in traditional venues. Consequently, using DBLP ensures a more comprehensive and timely collection of relevant literature for this study.

\subsection{Search String Formulation}

We developed the search string through a systematic and multi-stage process. First, the first three authors drew on their expertise in \toolname{} and software engineering to curate a set of seed papers that apply \toolname{} to software engineering tasks. We then analyzed the titles, abstracts, and keywords of these papers to extract an initial list of search terms. These terms were subsequently refined and expanded through brainstorming sessions to incorporate relevant synonyms and variations.

The final search string combines two core concepts: (1) terms related to the LLM-as-a-Judge paradigm and (2) terms related to the software engineering domain. The final query is:

\begin{center}
    \begin{tabular}{c}
    \texttt{((Agent OR LLM OR ``Large Language Model'') AND (Judg* OR Evaluat*))} \\
    \textbf{AND} \\
    \texttt{(Code OR Coding OR Program* OR Software OR API OR Bug OR Test)} \\
    \end{tabular}
\end{center}

Note that the asterisk (*) serves as a wildcard to capture morphological variations of a root word (e.g., \texttt{Evaluat*} matches ``evaluate,'' ``evaluator,'' and ``evaluation'').
Based on the selected keywords, we conduct 42 searches on DBLP
on August 7th, 2025.

\subsection{Snowballing}
Our initial keyword search on the DBLP database was limited to paper titles. To mitigate the risk of missing relevant studies with this approach and to ensure comprehensive coverage, we applied a snowballing methodology~\cite{jalali2012systematic, wohlin2014guidelines}. This involved backward snowballing by screening the reference lists of the selected papers and forward snowballing using Google Scholar to identify newer papers citing them. The procedure was conducted from August 7 to August 12, 2025, and was repeated until saturation was reached, with no additional relevant studies identified.

\subsection{Quality Assessment}
To mitigate potential biases introduced by low-quality studies, we established five Quality Assessment Criteria (QAC) tailored to the specific challenges of evaluating LLM-as-a-Judge systems in software engineering.

Among these, three QAC are applied to assess the quality of the selected studies. They examine each paper’s overall Research Contribution (QAC1), its Methodological Clarity and Reproducibility (QAC2), and its Experimental Rigor (QAC3). The complete quality assessment checklist is presented below.

\begin{description}
    \item[\textbf{QAC1: Research Contribution.}] 
    This criterion evaluates whether the study makes a meaningful contribution to the field. A contribution can be demonstrated by proposing a novel LLM-as-a-Judge framework for SE tasks, creating a high-quality benchmark, or conducting a rigorous empirical study that offers critical insights into the performance of LLM-as-a-Judge in SE.

    \item[\textbf{QAC2: Methodological Clarity and Reproducibility.}] 
    This criterion assesses the completeness and clarity of the paper's methodological description. The workflow, implementation details, and experimental procedures must be documented in sufficient detail to allow another researcher to faithfully replicate the study.

    \item[\textbf{QAC3: Experimental Rigor.}]
    This criterion evaluates the scientific soundness of the experimental design itself. It assesses whether the chosen datasets, baselines, and evaluation metrics are appropriate and well-justified for the research question, ensuring the validity of the study's conclusions.
\end{description}

To ensure reliability, the quality assessment checklist was independently applied to all primary studies by the first three authors. Any discrepancies were then resolved through discussion to reach a final consensus.

\subsection{Results}

\begin{figure}
    \centering
    \includegraphics[width=0.9\textwidth]{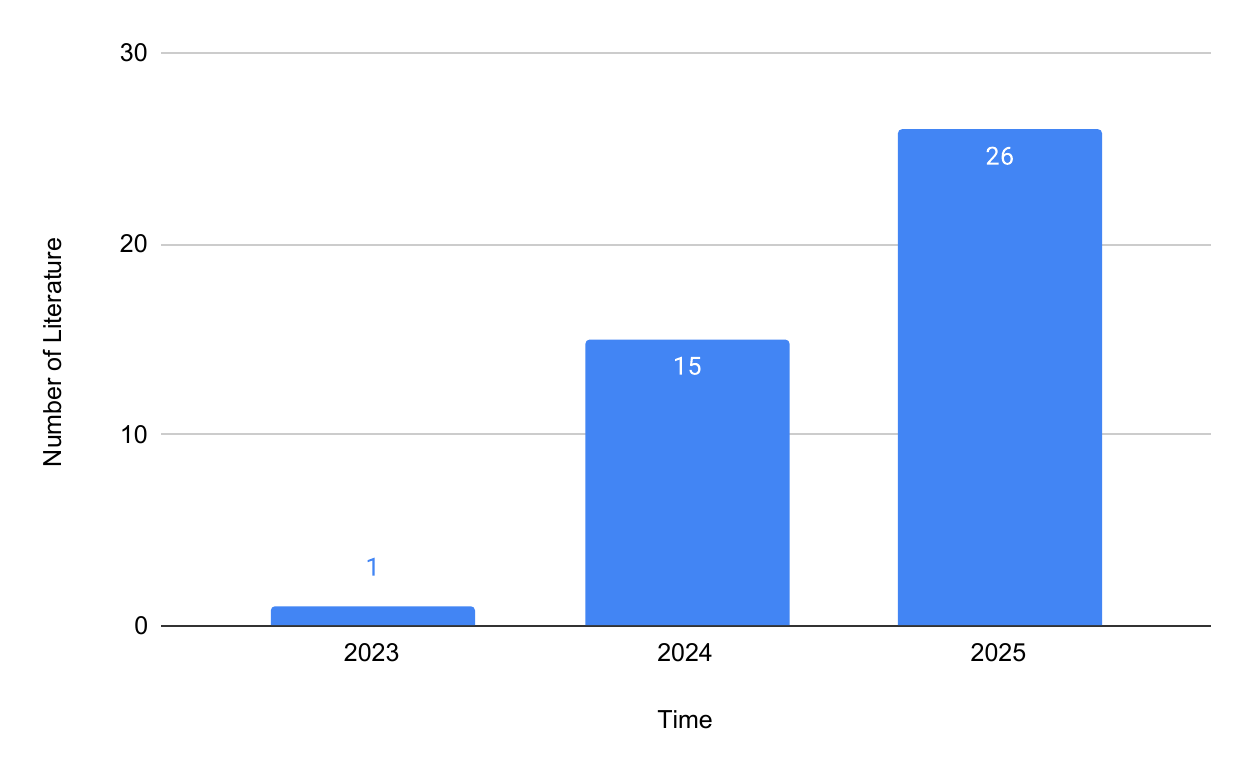}
    \caption{Annual Distribution of Publications on LLM-as-a-Judge in SE (2023 - Aug 2025)}
    \label{fig:literature}
\end{figure}

Our paper selection process yielded a total of 42 relevant publications. Figure \ref{fig:literature} illustrates their temporal distribution over 2023, 2024, and 2025.
Following the first work published in 2023, the field expanded with 15 publications in 2024. This trend intensified remarkably in 2025, with 26 publications cataloged by August alone. This exponential growth strongly indicates a surging research interest in the LLM-as-a-Judge paradigm within Software Engineering. The fact that the publication count for 2025 has already far surpassed the previous year's total in just eight months underscores the exceptional pace of development in this domain. For published works, the formal acceptance year was used as the publication date, which may differ from their initial posting on preprint servers like arXiv.

\section{Literature Review}
\label{sec:review}

In this section, we systematically review the application of the LLM-as-a-Judge paradigm across a spectrum of software engineering tasks. To provide a structured overview, we categorize the existing literature based on the SE domain being addressed. Our analysis covers four areas: \textbf{requirements engineering}, \textbf{coding assistance}, \textbf{software maintenance}, and \textbf{quality assurance}. These categories were chosen because they organically emerged from the reviewed literature and correspond to the fundamental stages of the software development lifecycle.

\begin{table}[]
    \centering
    \caption{A Taxonomy of LLM-as-a-Judge Literature in Software Engineering. The number of \toolname publications for evaluating each kind of SE artifacts is provided in brackets. A single paper may be cited for multiple SE artifacts.}
    \label{tab:literature}
    \resizebox{0.9\textwidth}{!}{%
    \begin{tabular}{lll}
    \hline
    \textbf{SE Activity}               & \textbf{SE Artifact}                      & \textbf{Reference}                                                                                                                                                          \\ \hline
    \multirow{3}{*}{Requirement Engineering} & Requirements Documents (2)               & \cite{lubos2024leveraging, quattrocchi2025can, ahmed2024can}                                                                                                                                  \\
                                       & System Specification (1)           & \cite{reinpold2024exploring}                                                                                                                                                   \\
                                       & User Stories (1) & \cite{quattrocchi2025can}                                                                                                                                                            \\ \hline
    \multirow{6}{*}{Coding Assistance} & Generated Code (27)                  & \cite{zhuge2024agent, zhou2025evaluating, yang2025code, koutcheme2025evaluating, wang2025mcts, simoes2024evaluating, li2025llms, vo2025llm, kim2025reproduction}            \\
                                       &                                       & \cite{vu2407foundational, wang2025codevisionary, zhou2025llm, wang2025can, tong-zhang-2024-codejudge, ahmed2024can, moon2025don, patel2024aime, tan2024judgebench, jaoua2025combining}          \\
                                       &                                       & \cite{zhao-etal-2025-codejudge, zhuo2023ice, jiang2025codejudgebench, weyssow2024codeultrafeedback, xu2024human, crupi2025effectiveness, findeis2025can, pan2025benchmarks} \\ \cline{2-3}
                                       & Code Summary (8)                & \cite{wang2025can, ahmed2024can, farchi2024automatic, wu2024can, zhou2025llm, mu2025evaluate, li2025llms, crupi2025effectiveness}             \\
                                       & Translated Code (2)                  & \cite{wang2025can, li2025llms}                                                                                                                                              \\
                                       & Answers to Software Questions  (2)       & \cite{zheng2024judging, xie2025prompting}                                                                                                                                   \\ \hline
    \multirow{3}{*}{Quality Assurance} & Bug Report (1)          & \cite{kumar2024llms}                                                                                                                                                        \\
                                       & Tests (3)             & \cite{wang2025rum, sollenberger2024llm4vv, jiang2025codejudgebench}                                                                                                                                  \\
                                       & Vulnerability Explanation (1)           & \cite{ weyssow2025r2vul}                                                                                                                                     \\ \hline
    \multirow{2}{*}{Maintenance}       & Commit Message (1)         & \cite{zeng2025evaluating}                                                                                                                                                   \\
                                       & Code Patches (5)           & \cite{yadavally2025large, ahmed2024can, li2024cleanvul, zhou2025llm, jiang2025codejudgebench}                                                                                                                     \\ \hline
    \end{tabular}%
    }
    \end{table}

\subsection{Requirements Engineering}

    The LLM-as-a-Judge paradigm is being effectively applied to the foundational phase of software development: requirements engineering. Approaches use LLMs to automate the evaluation of early-stage artifacts, ensuring the quality of documents before development begins. From Table \ref{tab:re}, we can see this involves documents ranging from user stories to technical system specifications.

    \begin{table}[]
\centering
\caption{Overview on the application of \toolname in Requirements Engineering}
\label{tab:re}
\resizebox{\textwidth}{!}{%
\begin{tabular}{ccccc}
\hline
                                                                     & \textbf{Evaluation Type} & \textbf{Evaluation Criteria}                                                                                                                                                               & \textbf{Evaluation Item}                                          & \textbf{Evaluation Result} \\ \hline
Ahmed et al.~\cite{ahmed2024can}                                     & Point-wise               & Presence of a causal relationship                                                                                                                                                          & \begin{tabular}[c]{@{}c@{}}Requirements \\ documents\end{tabular} & Graded                     \\
\rowcolor[HTML]{EFEFEF} Quattrocchi et al.~\cite{quattrocchi2025can} & Point-wise               & \begin{tabular}[c]{@{}c@{}}Feature Specificity, Rationale Clarity, \\ Problem-Oriented, Language Clarity,\\ Internal Consistency\end{tabular}                                              & User Stories                                                      & Graded                     \\
Lubos et al.~\cite{lubos2024leveraging}                              & Point-wise               & \begin{tabular}[c]{@{}c@{}}ISO-29148 quality characteristics \\ (Appropriate, Complete, \\ Conforming, Correct, Feasible, \\ Necessary, Singular, \\ Unambiguous, Verifiable)\end{tabular} & \begin{tabular}[c]{@{}c@{}}Requirements \\ documents\end{tabular} & Graded                     \\
\rowcolor[HTML]{EFEFEF} Reinpold et al.~\cite{reinpold2024exploring} & Point-wise               & \begin{tabular}[c]{@{}c@{}}Fulfillment and Applicability \\ of a requirement \\ with respect to the specification\end{tabular}                                                             & \begin{tabular}[c]{@{}c@{}}System \\ specification\end{tabular}   & Graded                     \\ \hline
\end{tabular}%
}
\end{table}

    \subsubsection{Requirements Documents}
    The \toolname paradigm is used to assess the quality of requirement artifacts. For instance, Lubos et al.~\cite{lubos2024leveraging} use Llama 2 (70B) to evaluate software requirements against the ISO 29148 standard\footnote{\url{https://www.iso.org/standard/72089.html}} for nine quality characteristics: Appropriate, Complete, Conforming, Correct, Feasible, Necessary, Singular, Unambiguous, and Verifiable. They find that the LLM not only identifies most quality flaws but also provides reliable explanations for them. 
    Ahmed et al.~\cite{ahmed2024can} demonstrate a more granular application of the \toolname paradigm by using an LLM to extract causal relationships from natural language requirements. The LLM provides a binary judgment on whether a sentence contains a causal link (e.g., 'If event 1, then event 2').

    \subsubsection{System Specifications}
    Reinpold et al.~\cite{reinpold2024exploring} apply LLMs to verify technical system specifications. Their work demonstrates that models like GPT-4o and Claude 3.5 Sonnet can effectively assess whether a system specification fulfills its corresponding requirements.

    \subsubsection{User Stories}
    Quattrocchi et al.~\cite{quattrocchi2025can} employ ten LLMs to assess the semantic quality of user stories against five criteria: Feature Specificity (measuring how precisely a software feature is described), Rationale Clarity (the clarity of the feature's justification), Problem-Oriented (evaluating whether the user story specifies a problem without implying a solution), Language Clarity (the clarity and precision of the language), and Internal Consistency (checking for contradictions).

    \subsection{Coding Assistant}
    This sub-section details its application across four primary coding assistance tasks: code generation, code summarization, code translation, and software question answering. Correspondingly, we focus on the evaluation of generated code, code summary, translated code, and answers to software questions.
    Table \ref{tab:code} provides an overview of the application of \toolname in Coding Assistance.

\begin{table}[]
\centering
\caption{Overview on the application of \toolname in Coding Assistance }
\label{tab:code}
\resizebox{0.8\textwidth}{!}{%
\begin{tabular}{ccccc}
\hline
\textbf{Reference}                                    & \textbf{Evaluation Type}                                         & \textbf{Evaluation Criteria}                                                                                                & \textbf{Evaluation Items}                                                                      & \textbf{Evaluation Result}                                                       \\ \hline
Agent-as-a-Judge~\cite{zhuge2024agent}                & Point-wise                                                       & \begin{tabular}[c]{@{}c@{}}Requirement \\ Satisfaction Rate\end{tabular}                                                    & Generated Code                                                                                 & Graded                                                                           \\
\rowcolor[HTML]{EFEFEF} 
Ahmed et al.~\cite{ahmed2024can}                      & Point-wise                                                       & \begin{tabular}[c]{@{}c@{}}Accuracy, Adequacy, \\ Conciseness, Similarity\end{tabular}                                      & \begin{tabular}[c]{@{}c@{}}Generated Code,\\ Code summary\end{tabular}                         & Graded                                                                           \\
AIME~\cite{patel2024aime}                             & Point-wise                                                       & \begin{tabular}[c]{@{}c@{}}Logic, Syntax,\\  Readability, Efficiency,\\ Redundancy\end{tabular}                             & Generated Code                                                                                 & Graded                                                                           \\
\rowcolor[HTML]{EFEFEF} 
Findeis et al.~\cite{findeis2025can}                  & Pair-wise,                                                       & Correctness                                                                                                                 & Generated Code                                                                                 & Best-Choice                                                                      \\
CODE-DITING~\cite{yang2025code}                       & Point-wise                                                       & Correctness                                                                                                                 & Generated Code                                                                                 & Graded                                                                           \\
\rowcolor[HTML]{EFEFEF} 
CodeJudge~\cite{tong-zhang-2024-codejudge}            & Point-wise                                                       & Correctness                                                                                                                 & Generated Code                                                                                 & Graded                                                                           \\
CodeJudge-Eval~\cite{zhao-etal-2025-codejudge}        & Point-wise                                                       & Correctness                                                                                                                 & Generated Code                                                                                 & Graded                                                                           \\
\rowcolor[HTML]{EFEFEF} 
CodeJudgeBench~\cite{jiang2025codejudgebench}         & \begin{tabular}[c]{@{}c@{}}Point-wise,\\ Pair-wise\end{tabular}  & Correctness                                                                                                                 & Generated Code                                                                                 & \begin{tabular}[c]{@{}c@{}}Best-Choice, \\ Graded\end{tabular}                   \\
Codeultrafeedback~\cite{weyssow2024codeultrafeedback} & \begin{tabular}[c]{@{}c@{}}Point-wise, \\ List-wise\end{tabular} & \begin{tabular}[c]{@{}c@{}}Coding preferences, \\ Correctness,\\ Usefulness\end{tabular}                                    & Generated Code                                                                                 & \begin{tabular}[c]{@{}c@{}}Graded,\\ Ranked-Ordering\end{tabular}                \\
\rowcolor[HTML]{EFEFEF} 
CodeVisionary~\cite{wang2025codevisionary}            & Point-wise                                                       & \begin{tabular}[c]{@{}c@{}}Code Correctness, \\ Functionality, Clarity\end{tabular}                                         & Generated Code                                                                                 & Graded                                                                           \\
Jaoua et al.~\cite{jaoua2025combining}                & \begin{tabular}[c]{@{}c@{}}Point-wise, \\ List-wise\end{tabular} & \begin{tabular}[c]{@{}c@{}}Relevance, \\ Accuracy, Coverage\end{tabular}                                                    & Generated Code                                                                                 & \begin{tabular}[c]{@{}c@{}}Graded, \\ Rank Ordering\end{tabular}                 \\
\rowcolor[HTML]{EFEFEF} 
Crupi et al.~\cite{crupi2025effectiveness}            & Point-wise                                                       & \begin{tabular}[c]{@{}c@{}}Correctness, Adequacy, \\ Conciseness, Fluency, \\ Understandability\end{tabular}                & \begin{tabular}[c]{@{}c@{}}Generated Code,\\ Code Summary\end{tabular}                         & Graded                                                                           \\
Moon et al.~\cite{moon2025don}                        & Point-wise                                                       & Correctness                                                                                                                 & Generated Code                                                                                 & Graded                                                                           \\
\rowcolor[HTML]{EFEFEF} 
JETTS~\cite{zhou2025evaluating}                       & \begin{tabular}[c]{@{}c@{}}Point-wise, \\ List-wise\end{tabular} & \begin{tabular}[c]{@{}c@{}}Correctness,\\ Overall Quality\end{tabular}                                                      & Generated Code                                                                                 & \begin{tabular}[c]{@{}c@{}}Graded, \\ Best-Choice, \\ Rank Ordering\end{tabular} \\
Koutcheme et al.~\cite{koutcheme2025evaluating}       & Point-wise                                                       & \begin{tabular}[c]{@{}c@{}}Accuracy, \\ Selectivity, \\ Clarity\end{tabular}                                                & Generated Code                                                                                 & Graded                                                                           \\
\rowcolor[HTML]{EFEFEF} 
Simõeset al.~\cite{simoes2024evaluating}              & Point-wise                                                       & \begin{tabular}[c]{@{}c@{}}Readability, \\ Overall Quality\end{tabular}                                                     & Java class files                                                                               & Graded                                                                           \\
Farchi et al.~\cite{farchi2024automatic}              & Pair-wise                                                        & \begin{tabular}[c]{@{}c@{}}Usefulness, \\ Similarity,\\ Consistency\end{tabular}                                            & Code summaries                                                                                 & Graded                                                                           \\
\rowcolor[HTML]{EFEFEF} 
Vu et al.~\cite{vu2407foundational}                   & \begin{tabular}[c]{@{}c@{}}Pair-wise, \\ List-wise\end{tabular}  & \begin{tabular}[c]{@{}c@{}}Correctness,\\ Quality\end{tabular}                                                              & Generated Code                                                                                 & Best-Choice                                                                      \\
Kim et al.~\cite{kim2025reproduction}                 & List-wise                                                        & Correctness                                                                                                                 & Generated Code                                                                                 & Best-Choice                                                                      \\
\rowcolor[HTML]{EFEFEF} 
HuCoSC~\cite{xu2024human}                             & Pair-wise                                                        & \begin{tabular}[c]{@{}c@{}}Logical correctness, \\ Execution correctness, \\ Efficiency.\end{tabular}                       & Generated Code                                                                                 & Graded                                                                           \\
ICE-Score~\cite{zhuo2023ice}                          & Point-wise                                                       & \begin{tabular}[c]{@{}c@{}}Usefulness, \\ Correctness\end{tabular}                                                          & Generated Code                                                                                 & Graded                                                                           \\
\rowcolor[HTML]{EFEFEF} 
Judgebench~\cite{tan2024judgebench}                   & Pair-wise                                                        & \begin{tabular}[c]{@{}c@{}}Factual Correctness,\\  Logical Correctness\end{tabular}                                         & Generated Code                                                                                 & Best-Choice                                                                      \\
Li et al.~\cite{li2025llms}                           & \begin{tabular}[c]{@{}c@{}}Point-wise, \\ Pair-wise\end{tabular} & \begin{tabular}[c]{@{}c@{}}Helpfulness, \\ Functional correctness, \\ syntactic validity, \\ semantic accuracy\end{tabular} & \begin{tabular}[c]{@{}c@{}}Generated Code,\\ Code Summary,\\ Translated Code\end{tabular}      & \begin{tabular}[c]{@{}c@{}}Graded, \\ Best-Choice\end{tabular}                   \\
\rowcolor[HTML]{EFEFEF} 
Vo et al.~\cite{vo2025llm}                            & \begin{tabular}[c]{@{}c@{}}Point-wise, \\ Pair-wise\end{tabular} & \begin{tabular}[c]{@{}c@{}}Functionality matching, \\ Logic correctness\end{tabular}                                        & Generated Code                                                                                 & Graded                                                                           \\
MCTS-Judge~\cite{wang2025mcts}                        & Point-wise                                                       & \begin{tabular}[c]{@{}c@{}}Code logic, \\ Completeness, \\ Correctness\end{tabular}                                         & Generated Code                                                                                 & \begin{tabular}[c]{@{}c@{}}Graded, \\ Best-Choice\end{tabular}                   \\
\rowcolor[HTML]{EFEFEF} 
Mu et al.~\cite{mu2025evaluate}                       & Pair-wise                                                        & \begin{tabular}[c]{@{}c@{}}Natural Semantics, \\ Readability, \\ Naturalness\end{tabular}                                   & \begin{tabular}[c]{@{}c@{}}Generated Code,\\ Code Summaries\end{tabular}                       & Graded                                                                           \\
Wang et al.~\cite{wang2025can}                        & \begin{tabular}[c]{@{}c@{}}Point-wise\\ Pair-wise,\end{tabular}  & \begin{tabular}[c]{@{}c@{}}Functional correctness, \\ Consistency, \\ Readability, \\ Idiomatic Usage\end{tabular}          & \begin{tabular}[c]{@{}c@{}}Generated Code,\\ Code summary,\\ Translated Code\end{tabular}      & \begin{tabular}[c]{@{}c@{}}Graded\\ Best-Choice,\end{tabular}                    \\
\rowcolor[HTML]{EFEFEF} 
Pan et al.~\cite{pan2025benchmarks}                   & Point-wise                                                       & \begin{tabular}[c]{@{}c@{}}Correctness, \\ Adherence to \\ specification\end{tabular}                                       & Generated Code                                                                                 & Graded                                                                           \\
CODERPE~\cite{wu2024can}                              & Point-wise                                                       & \begin{tabular}[c]{@{}c@{}}Coherence, \\ Consistency,\\  Fluency, \\ Relevance\end{tabular}                                 & Code summary                                                                                   & Graded                                                                           \\
\rowcolor[HTML]{EFEFEF} 
Xie et al.~\cite{xie2025prompting}                    & Point-wise                                                       & \begin{tabular}[c]{@{}c@{}}Factual Content \\ Semantic Similarity, \\ Fact Match Ratio\end{tabular}                         & \begin{tabular}[c]{@{}c@{}}Responses to queries about \\ Databricks documentation\end{tabular} & Graded                                                                           \\
Zheng et al.~\cite{zheng2024judging}                  & \begin{tabular}[c]{@{}c@{}}Point-wise, \\ Pair-wise\end{tabular} & \begin{tabular}[c]{@{}c@{}}Helpfulness, Relevance, \\ Accuracy, Depth, \\ Creativity, Details\end{tabular}                  & \begin{tabular}[c]{@{}c@{}}Answers to \\ coding questions\end{tabular}                         & \begin{tabular}[c]{@{}c@{}}Graded, \\ Best-Choice\end{tabular}                   \\
\rowcolor[HTML]{EFEFEF} 
Zhou et al.~\cite{zhou2025llm}                        & \begin{tabular}[c]{@{}c@{}}Point-wise,\\ Pair-wise\end{tabular}  & \begin{tabular}[c]{@{}c@{}}Correctness,\\ Semantic \\ Equivalence\end{tabular}                                              & \begin{tabular}[c]{@{}c@{}}Generated Code,\\ Code summary\end{tabular}                         & Graded                                                                           \\ \hline
\end{tabular}%
}
\end{table}

    \subsubsection{Generated Code.}
    Effectively and automatically evaluating the quality of generated code is one of the most critical problems in software engineering.
    The evaluation of generated code is a primary application for LLM-as-a-Judge. Before the introduction of \toolname, automated code evaluation primarily relied on execution-based method, i.e., executing the code based on pre-defined test cases~\cite{zhuo2024bigcodebench, chen2021evaluating, hendrycks2021measuring}. 
    However, passing all the test cases does not necessarily mean the quality of the code is good, as the test cases may not comprehensively cover all the edge cases~\cite{dou2024s}.
    Constructing test cases can also be particularly challenging for certain tasks~\cite{tong-zhang-2024-codejudge}, such as AI model training, web scraping, etc. 
    When test cases are unavailable, many studies rely on reference-based metrics~\cite{dehaerne2022code, zheng2023codegeex}, such as CodeBLEU~\cite{ren2020codebleu} and CodeBERTScore~\cite{zhou2023codebertscore}. Nevertheless, the quality of the reference code restricts the effectiveness of reference-based metrics. Prior research~\cite{evtikhiev2023out} highlighted that reference-based metrics frequently misalign human judgments in code generation. 
    
    Consequently, a substantial body of research has explored the effectiveness of LLMs in evaluating code generation~\cite{wang2025can, zheng2024judging, tong-zhang-2024-codejudge, patel2024aime, tan2024judgebench, zhao-etal-2025-codejudge, zhuo2023ice, yang2025code, zhou2025evaluating}. We summarize three main characteristics of \toolname that are different from the execution-based and reference-based metrics:
    
    \begin{itemize}[leftmargin=*]
        \item\textit{Execution-free:} It does not require compulsory execution of the code.
        \item\textit{Reference-free:} Unlike reference-based metrics, \toolname does not require a reference code. Even though we can still optionally provide the reference code. 
        \item\textit{Multi-Facet Evaluation:} \toolname assesses intrinsic code qualities that traditionally required human judgment, such as readability and usefulness. Further, it can also provide qualitative, actionable feedback that explains the rationale behind its assessment.
    \end{itemize}

    \noindent \textit{\textbf{Common Evaluation Criteria.}} We summarize the common evaluation criteria used in the literature for code generation. We categorize them into two main criteria: \textit{Code Functionality} and \textit{Code Quality}. 
    
    \noindent\textit{Code Functionality:} This dimension assesses whether the code behaves correctly according to the specified requirements. It focuses on the  correctness of the code's output and behavior. Key criteria include:

    \begin{itemize}[leftmargin=*, label=-]
       \item \textbf{Functional Correctness:} Verifies whether the code produces the expected output according to the task description.
       \item \textbf{Fault Tolerance \& Error Handling:} Evaluates how the code manages edge cases, exceptions, invalid inputs, and unexpected failures.
    \end{itemize}
      
    \noindent \textit{Code Quality:} These criteria evaluate intrinsic attributes such as readability and adherence to best practices. Key criteria include:
    \begin{itemize}[leftmargin=*, label=-]
        \item \textbf{Complexity \& Efficiency:} Analyzes computational complexity, execution time, and memory usage.
        \item \textbf{Helpfulness:} Evaluates whether the code contributes meaningfully to solving the problem. This includes distinguishing partially correct solutions from completely unuseful and irrelevant code.
        \item \textbf{Readability:} Measures how easily the code can be understood, including aspects like fluency, clarity, and conciseness.
        \item \textbf{Stylistic Consistency:} Ensures adherence to established coding standards, formatting guidelines, and best practices.
        \item \textbf{Reference Similarity:} Assesses how closely the generated code aligns with a given reference implementation.
        \item \textbf{Minor Anomalies \& Warnings:} Identifies non-critical issues such as redundant code or unused variables that do not affect execution.
    \end{itemize}

    While promising, recent studies highlight significant challenges for \toolname in evaluating code generation. An extensive empirical study by Crupi et al.~\cite{crupi2025effectiveness} found that even powerful models like GPT-4-turbo frequently misjudge code correctness. This limitation has driven research beyond simple prompting techniques~\cite{zhuo2023ice,weyssow2024codeultrafeedback,crupi2025effectiveness} toward more sophisticated frameworks.
    
    One set of methods focus on utilizing prompt engineering to instruct \toolname to perform an evaluation. The common methods are summarized below.

    \begin{itemize}[leftmargin=*, label=$\circ$]
        \item \textit{Structured Prompting.} This is the foundational method where the LLM is given a detailed, structured prompt that clearly defines the task, the evaluation criteria, the scoring scale, and the required output format. ICE-Score~\cite{zhuo2023ice} is a prime example of this approach, instructing the LLM to follow specific steps to evaluate code, which proved more reliable than simple and direct prompts. CodeJudge~\cite{tong-zhang-2024-codejudge} leverages a detailed taxonomy of common programming errors in the prompt to guide the LLM in analyzing the generated code. 
        \item \textit{Decomposition and Multi-Step Reasoning.}
        To move beyond superficial analysis, researchers have developed methods that prompt the LLM to engage in a more deliberate, multi-step reasoning process~\cite{xu2024human, vo2025llm, wang2025mcts}. One key strategy involves simplifying the code itself. This  often achieved by first abstracting the code into a higher-level representation before judgment. For example, HuCoSC~\cite{xu2024human} mimics human cognition by performing Recursive Semantic Comprehension. It recursively decomposes code via its Abstract Syntax Tree (AST) into smaller parts. Vo et al.~\cite{vo2025llm} first translate the code into pseudocode. This allows the LLM judge to assess the core algorithm without being confused by complex, language-specific syntax. Further, another approach is to decompose the evaluation process. MCTS-Judge~\cite{wang2025mcts} reframes evaluation as a search problem, where the goal is to find the most reliable reasoning path. It uses Monte Carlo Tree Search~\cite{browne2012survey} to explore a tree of possible reasoning trajectories, where each path represents a unique sequence of evaluation sub-tasks (e.g., analyzing logic, then checking functionality).
    \end{itemize}

    Another set of frameworks~\cite{zhuge2024agent, wang2025codevisionary, findeis2025can} gives the LLM judge more autonomy (\textbf{Agentic}) and access to external information and tools. The Agent-as-a-Judge framework~\cite{zhuge2024agent} utilizes an agentic system with modules for file system interaction, dependency graphing, and code execution. This design allows the agent to interact with the output of another agent and provide detailed, step-by-step feedback. Similarly, CodeVisionary~\cite{wang2025codevisionary} employs an agent with a suite of tools, including dynamic code execution, static linting, and unit testing. Findeis et al.~\cite{findeis2025can} evaluate LLM's ability to distinguish between correct and incorrect Python solutions. They demonstrate that augmenting the LLM judge with a code execution tool increased its agreement with ground truth from below 42\% to approximately 72\%.

    To enhance reliability and mitigate the biases of a single LLM, methods are proposed to leverage the power of multiple evaluators and then employ various strategies to combine judgments~\cite{zhou2025llm, patel2024aime, wang2025codevisionary}. For instance, the SWE-Judge framework~\cite{zhou2025llm} aggregates evaluations scores from a dynamically assembled team of judges. Patel et al. propose AIME framework~\cite{patel2024aime}, which uses a specialist approach, assigning different LLMs to evaluate distinct criteria such as correctness and readability. They then theoretically prove the benefit of multiple evaluators over a single LLM evaluator. Other systems, like CodeVisionary~\cite{wang2025codevisionary}, facilitate direct negotiation among multiple LLM judges to reach a final consensus on the evaluation score.
    
    Another research direction focuses on creating smaller, more efficient, and specialized judges~\cite{yang2025code, vu2407foundational}. These approaches typically aim to improve performance, reduce computational costs, and increase explainability. Key strategies include:

    \begin{itemize}[leftmargin=*, label=$\circ$]
        \item \textit{Knowledge Distillation}: This method transfers capabilities from a large model to a smaller one. For instance, the CODE-DITING framework~\cite{yang2025code} uses a reasoning-focused data distillation process to create powerful 1.5B and 7B parameter judges from a 671B reasoning model (DeepSeek-R1). The resulting compact models are highly efficient, with the 7B version surpassing the performance of much larger models like GPT-4o.
        \item \textit{Fine-Tuning on Human Judgments}: This method uses a large dataset of human judgments to fine-tune the LLM Judge model. For instance, Vu et al.~\cite{vu2407foundational} curated the FLAMe collection, a dataset with over 5.3 million human judgments across 102 diverse quality assessment tasks, which includes specialized datasets for coding that teach the model to distinguish between correct and buggy programs based on coding problems or GitHub issues. They then fine-tuned a PaLM-2-24B model on this collection to create the FLAMe autorater models.
    \end{itemize}

\noindent \textit{\textbf{Improving Code Generation via LLM-as-a-Judge.}} LLM-as-a-Judge can be actively used to improve the output of code generation models. The most common applications are 1. Selection from Multiple Solutions (Verification \& Reranking) and 2. Iterative Refinement with Feedback.

\begin{itemize}[leftmargin=*, label=$\circ$]
    \item \textit{Selection from Multiple Solutions.} Kim et al.~\cite{kim2025reproduction} used an LLM judge as a verifier to select the best solution from multiple candidates, as code generation models often exhibit a large gap between single-shot success (Pass@1) and multi-shot success (Pass@5). Kim et al. utilizes \toolname to close this gap without requiring a ground-truth test suite. Vu et al.~\cite{vu2407foundational} trained the FLAMe autorater models on a dataset of over 5.3 million human judgments across 102 quality assessment tasks, including specialized coding datasets that teach the model to distinguish correct from buggy programs using coding problems or GitHub issues. They used FLAMe to improve code generation on the HumanEval Python benchmark. In this experiment, FLAMe re-ranked 10 code samples generated by other models. Selecting the top-ranked sample significantly improved Pass@1. For example, it raised CodeGen-16B’s accuracy from 21.2\% to 31.1\%.
    \item \textit{Iterative Refinement with Feedback.} Another directions uses the judge's output as feedback to iteratively refine the code generation model.
    Research shows that feedback quality is critical. Zhou et al.~\cite{zhou2025evaluating} found that generic, unstructured natural-language critiques from LLM judges were largely ineffective for enabling effective code refinement.
    Vo et al.~\cite{vo2025llm} built a Reflection code agent that used feedback to automatically fix incorrect Bash scripts.
    Weyssow et al.~\cite{weyssow2024codeultrafeedback} used the feedback from the LLM judge to align CodeLlama-7B with coding preferences. Experiments showed improved adherence to coding style and higher functional correctness.
\end{itemize}
    
\subsubsection{Code Summary.}

The \toolname paradigm is particularly well-suited for evaluating code summaries. Unlike code generation, where functional correctness can be verified through execution, the quality of a summary is subjective and relies on human perception. Consequently, LLM judges show strong potential in this domain, with multiple studies confirming that their evaluations align more closely with human judgment than traditional metrics~\cite{zhou2025llm, ahmed2024can, wu2024can, crupi2025effectiveness}.

For instance, Crupi et al.~\cite{crupi2025effectiveness} found that GPT-4 achieves moderate to high agreement with human evaluators, especially when assessing content adequacy. To further enhance evaluation quality, innovative prompting strategies have been developed. CODERPE~\cite{wu2024can}, for example, instructs the LLM judge to adopt different professional personas—such as a code reviewer, author, or system analyst—to assess summaries from multiple perspectives.

Common criteria for evaluating code summaries include:
\begin{itemize}[leftmargin=*]
    \item \textbf{Language-related:} Ensure clear, well-structured sentences with precise and appropriate wording.
    \item \textbf{Content-related:} Summarize the core functionality and logic concisely, providing sufficient detail without including unnecessary information.
    \item \textbf{Effectiveness-related:} Evaluate whether the summary is useful for developers and enhances their understanding of the code.
\end{itemize}

\subsubsection{Translated Code.}

Code translation involves converting code from one programming language to another while preserving its functionality~\cite{yan2310codetransocean}. 
Empirical evidence~\cite{wang2025can} suggests that the \toolname paradigm is particularly effective for this evaluation task.
Wang et al.~\cite{wang2025can} focused on translations involving, Java, Python, C, and C++. They found that \toolname reached a Pearson correlation of 0.81 with human scores on the CodeTransOcean~\cite{yan2310codetransocean} dataset, whereas the best traditional metric ChrF++~\cite{popovic2017chrf++}, only achieved 0.34. The success of \toolname is mainly attributed to the ability of LLMs to assess functional and semantic equivalence. Since traditional metrics rely on lexical matching, they often penalize correct translations that are syntactically different from a reference solution. In contrast, LLMs can understand the underlying program logic, enabling a more robust and reference-free evaluation.

\subsubsection{Answers to Software Question.}
Zheng et al.~\cite{zheng2024judging} investigated using LLMs like GPT-4 to judge responses to complex, open-ended queries, including coding and reasoning tasks. They found that GPT-4's judgments align remarkably well with human preferences, achieving over 80\% agreement, which is comparable to inter-human agreement. Xie et al.~\cite{xie2025prompting} applied \toolname to evaluate answers to questions about Databricks documentation. Their work addresses a key limitation where LLM judges often overemphasize minor details while undervaluing critical information. They introduced a prompting mechanism that instructs the judge to differentiate between ``Critical Facts,'' ``Supporting Facts,'' and ``Trivial Facts.'' This weighting system allows the judge to prioritize essential information and avoid penalizing answers for minor omissions, improving the human alignment rate by an average of 6\%.

\subsection{Quality Assurance} 
In software engineering, Quality Assurance (QA) encompasses a broad range of activities aimed at ensuring software quality, from identifying bugs to validating test coverage and assessing security. The \toolname paradigm is being applied to automate the evaluation of several key QA artifacts.

\begin{table}
    \centering
    \caption{Overview on the application of \toolname in Quality Assurance}
    \label{tab:qa}
    \resizebox{\linewidth}{!}{%
    \begin{tabular}{ccccc} 
    \hline
                                                                                                      & \textbf{Evaluation Type} & \textbf{Evaluation Criteria}                                                                                                                             & \textbf{Evaluation Item}                                                                                              & \textbf{Evaluation Result}                                                                         \\ 
    \hline
    RUM~\cite{wang2025rum}                                                                            & Point-wise               & \begin{tabular}[c]{@{}c@{}}Standardization, Sufficiency, \\ Consistency, Readability,\\Runnability\end{tabular}                                          & \begin{tabular}[c]{@{}c@{}}Test case document,\\Test script code,\\Test step screenshots\end{tabular}                 & Graded                                                                                             \\
    \rowcolor[rgb]{0.937,0.937,0.937} \cellcolor[HTML]{EFEFEF}LLMs as Evaluators~\cite{kumar2024llms} & List-wise                & \begin{tabular}[c]{@{}>{\cellcolor[rgb]{0.937,0.937,0.937}}c@{}}Factual Correctness, \\Completeness,\\No Hallucinations,\\Adequate Coverage\end{tabular} & \begin{tabular}[c]{@{}>{\cellcolor[rgb]{0.937,0.937,0.937}}c@{}}Bug report titles,\\Bug report summaries\end{tabular} & \begin{tabular}[c]{@{}>{\cellcolor[rgb]{0.937,0.937,0.937}}c@{}}Graded,\\Best-Choice\end{tabular}  \\
    LLM4VV~\cite{sollenberger2024llm4vv}                                                              & Point-wise               & \begin{tabular}[c]{@{}c@{}}Syntax, Directive Appropriateness, \\ Clause Correctness, \\Memory Management, \\ Compliance, Logic\end{tabular}              & Compiler test code                                                                                                    & Graded                                                                                             \\
    \rowcolor[rgb]{0.937,0.937,0.937} R2VUL~\cite{weyssow2025r2vul}                                   & Point-wise               & \begin{tabular}[c]{@{}>{\cellcolor[rgb]{0.937,0.937,0.937}}c@{}}Completeness, Clarity,\\ Actionability\end{tabular}                                      & \begin{tabular}[c]{@{}>{\cellcolor[rgb]{0.937,0.937,0.937}}c@{}}Vulnerbility \\Explanations\end{tabular}              & Graded                                                                                             \\
    \hline
    \end{tabular}
    }
    \end{table}

\subsubsection{Bug Report.}
For evaluating bug reports, LLM judges have shown significant promise. A study by Kumar et al.~\cite{kumar2024llms} found that advanced models like GPT-4o consistently matched or exceeded human accuracy in selecting the most accurate titles and summaries for bug reports. Crucially, the study noted that while human evaluators experienced fatigue that degraded their accuracy over time, the LLM’s performance remained stable. This consistency makes \toolname a scalable and potentially more reliable alternative for large-scale QA evaluations.

\subsubsection{Tests.}
Sollenberger et al.~\cite{sollenberger2024llm4vv} apply \toolname to evaluate the compiler test suite. These are tests designed to validate and verify compiler implementations for parallel programming models of OpenMP\footnote{\url{https://www.openmp.org/}} and OpenACC\footnote{\url{https://www.openacc.org/}}. They demonstrate an agent-based approach where the LLM judge is provided with compiler outputs and execution results to assess the correctness of compiler test suites.
Wang et al.~\cite{wang2025rum} propose RUM, a hybrid system that delegates simple, objective checks (e.g., syntax, naming conventions) to fast, rule-based systems, reserving the LLM for more subjective assessments like test case readability and coverage. Their evaluation covers a comprehensive set of testing work products, including Test Case Documents, Test Script Code, and Test Reports. This division optimizes for both cost and quality, achieving a high correlation with human expert scores.

\subsubsection{Vulnerability Explanation.}
Instead of making direct and high-stakes vulnerability judgments, as explored in prior studies~\cite{ZhouZ024,ZhouCSL25}, \toolname are increasingly employed in meta-evaluation roles. For example, recent work~\cite{weyssow2025r2vul} demonstrates that they can assess whether vulnerability explanations are logically consistent, sufficiently detailed, and clearly articulated. In this way, LLMs serve less as ultimate arbiters of security and more as scalable reviewers that enhance the reliability and interpretability of automated vulnerability analysis.

\subsection{Maintenance}

Software maintenance, which involves modifying and updating code after its initial release, is another critical area where the \toolname paradigm is applied. The focus here is on evaluating the quality of the artifacts that document and implement these changes, namely commit messages and code patches.

\subsubsection{Commit Message.}
The LLM-as-a-Judge paradigm has proven to be a effective method for evaluating the quality of automatically generated commit messages~\cite{zeng2025evaluating}. Zeng et al.~\cite{zeng2025evaluating} demonstrated that when using advanced prompting strategies combining Chain-of-Thought and few-shot examples, LLMs like GPT-4 can achieve evaluation performance comparable to human experts. The study assessed messages on their ability to explain both ``what'' a change does and ``why'' it was made. The LLM-based approach significantly outperformed traditional reference-based metrics (e.g., BLEU, ROUGE), effectively overcoming the problem where multiple, lexically different commit messages can be equally valid for a single code change. Crucially, the research also confirmed the stability of the LLM evaluator, showing acceptable levels of reproducibility and robustness.

\subsubsection{Code Patch.}

The \toolname paradigm is increasingly applied to the evaluation of code patches~\cite{zhou2025llm,li2024cleanvul,yadavally2025large,ahmed2024can}, a critical task in automated program repair and vulnerability mitigation. Yadavally et al.~\cite{yadavally2025large} achieve high accuracy by using an LLM to predict, in an execution-free manner, whether a given patch would pass unseen tests, while Ahmed et al.~\cite{ahmed2024can} employ LLMs to judge code patches aimed at fixing static analysis warnings, a task requiring deep reasoning. More recently, Li et al.~\cite{li2024cleanvul} introduce CleanVul, which leverages LLM heuristics to identify security-related patches and construct a high-quality dataset of them, showing that LLM-based judgments can significantly improve the correctness and generalizability of downstream vulnerability detectors. More broadly, Zhou et al.~\cite{zhou2025llm} propose SWE-Judge, an ensemble-based LLM-as-a-Judge metric that integrates multiple evaluation strategies (e.g., direct assessment, semantic equivalence checking, and test generation) to better align patch evaluations with human judgments. Collectively, these studies highlight the expanding role of LLM-as-a-Judge frameworks not only in patch-level correctness evaluation but also in dataset curation and the establishment of reliable evaluation protocols for automated program repair and vulnerability mitigation.

\begin{table}
\centering
\caption{Overview on the application of \toolname in Software Maintenance}
\label{tab:main}
\resizebox{0.9\linewidth}{!}{%
\begin{tabular}{ccccc} 
\hline
                                                                                                    & \textbf{Evaluation Type}                                                                            & \textbf{Evaluation Criteria}                                                                                                 & \textbf{Evaluation Item} & \textbf{Evaluation Result}  \\ 
\hline
Zeng et al.~\cite{zeng2025evaluating}                                                               & Point-wise                                                                                          & \begin{tabular}[c]{@{}c@{}}Content Quality \\(What \& Why)\end{tabular}                                                         & Commit Message           & Graded                      \\
\rowcolor[rgb]{0.937,0.937,0.937} \cellcolor[HTML]{EFEFEF}Yadavally et al~\cite{yadavally2025large} & Point-wise                                                                                          & \begin{tabular}[c]{@{}>{\cellcolor[rgb]{0.937,0.937,0.937}}c@{}}Functional Correctness,\\Functional\end{tabular}             & Code Patch               & Graded                      \\
Li et al.~\cite{li2024cleanvul}                                                                     & Point-wise                                                                                          & Vulnerability Relevance                                                                                                      & Code Patch               & Graded                      \\
\rowcolor[rgb]{0.937,0.937,0.937} \cellcolor[HTML]{EFEFEF}SWE-Judge~\cite{zhou2025llm}              & \begin{tabular}[c]{@{}>{\cellcolor[rgb]{0.937,0.937,0.937}}c@{}}Point-wise,\\Pair-wise\end{tabular} & \begin{tabular}[c]{@{}>{\cellcolor[rgb]{0.937,0.937,0.937}}c@{}}Functional Correctness,\\Functional Equivalence\end{tabular} & Code Patch               & Graded                      \\
Ahmed et al.~\cite{ahmed2024can}                                                                    & Point-wise                                                                                          & \begin{tabular}[c]{@{}c@{}}Actionability (whether fixes \\ the warning of static analyzers)\end{tabular}                     & Code Patch               & Graded                      \\
\hline
\end{tabular}
}
\end{table}

\section{The Road Ahead}
\label{sec:roadmap}

This section outlines a research roadmap for achieving scalable, reliable, and effective LLM-as-a-Judge systems in software engineering. 
We begin by outlining the key limitations in the current landscape and then propose research opportunities and concrete actionable steps that can significantly broaden the applicability of \toolname across diverse SE contexts, steering us toward our envisioned future.

\begin{figure}
    \centering
    \includegraphics[width=0.9\textwidth]{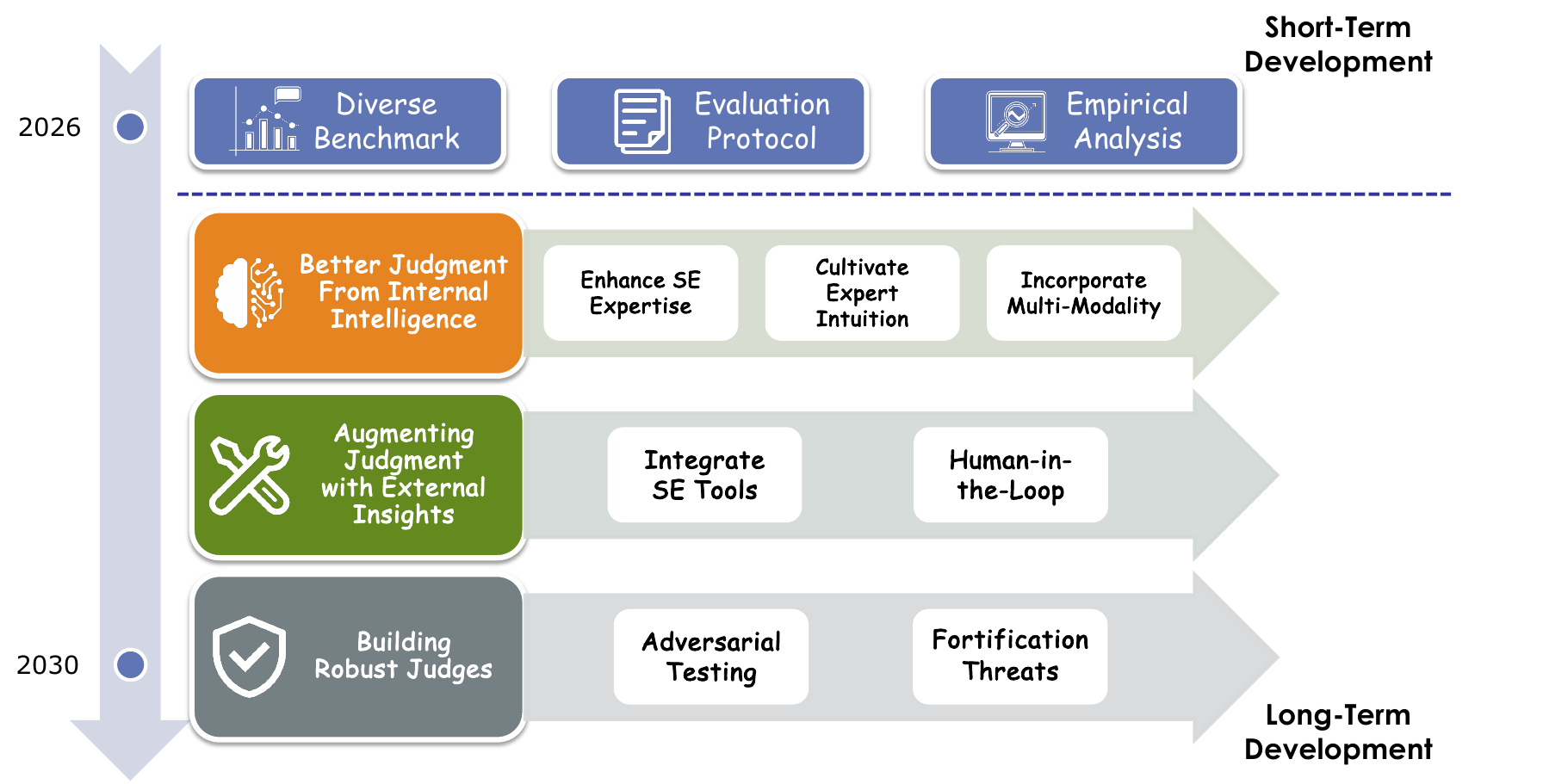}
    \caption{The Roadmap for LLM-as-a-Judge in SE.}
    \label{fig:roadmap}
\end{figure}

The proposed research roadmap, visualized in Figure \ref{fig:roadmap}, charts a course for advancing LLM-as-a-Judge systems in software engineering by 2030. The roadmap is structured around two key phases:

\begin{itemize}[leftmargin=*]
    \item \textbf{Foundational Short-Term Goals:} The initial phase focuses on building a stable foundation for future research.
    \begin{enumerate}
    \item \textbf{Develop Comprehensive Benchmarks and a Nuanced Evaluation Protocol:} This goal aims to overcome the scarcity of large-scale, human-annotated benchmarks and the limitations of current metrics. The objective is to create diverse benchmarks with high-quality expert annotations covering nuanced criteria like readability and helpfulness. The new protocol will account for subjectivity in SE tasks by capturing the full distribution of expert opinions and using distribution-aware metrics for assessment.
    \item \textbf{Conduct Comprehensive Empirical Evaluations and Bias Analysis:} This goal seeks to resolve inconsistent empirical findings in the field. The plan is to conduct large-scale, systematic evaluations of LLM-as-a-Judge systems across various SE tasks. These studies will analyze key design variables, such as prompting strategies and LLM configurations, and will include in-depth investigations into cognitive biases that cause flawed judgments.
    \end{enumerate}
    \item \textbf{Parallel Long-Term Tracks:} Building on this foundation, three parallel research tracks will advance LLM judges into sophisticated systems for the SE lifecycle.
    \begin{enumerate}
    \item \textbf{Enhancing Internal Intelligence:} Research will focus on improving the SE expertise of LLMs by training them on high-quality, domain-specific data and capturing the procedural knowledge of human experts. Another key advancement would be the integration of multi-modal assessment, enabling MLLM judges to evaluate visual artifacts like UML diagrams and GUI mockups.
    \item \textbf{Augmenting with External Insights:} This focus on creating frameworks that integrate the LLM judge with a diverse set of external SE tools, such as formal verification systems. It also aims to establish a synergistic relationship between LLM judges and human experts, creating a continuous feedback loop where human corrections are used to refine the LLM judge.
    \item \textbf{Building Robust and Secure Judges:} This track focuses on systematically testing for vulnerabilities and strengthening defenses against advanced adversarial threats. Research will move beyond simple attacks to uncover vulnerabilities through advanced adversarial testing, enabling the development of resilient judges fortified against sophisticated, semantics-preserving manipulations.
    \end{enumerate}
    \end{itemize}

\subsection{Developing Larger-Scale Benchmarks and Rigorous Evaluation Frameworks}

\noindent \textbf{Limitation 1: Lack of Large-Scale High-Quality Human-Annotated Benchmarks. }
In Section \ref{sec:review}, we reviewed numerous studies evaluating the effectiveness of LLM-as-a-Judge in software engineering~\cite{wang2025can,ahmed2024can,wang2025codevisionary,crupi2025effectiveness,zhou2025llm}. These literature typically assess the alignment between LLM-as-a-Judge and human judgment. However, a foundational limitation exists in the field is the scarcity of large-scale, human-annotated benchmarks for evaluating LLM-as-a-Judge systems. While established benchmarks such as HumanEval~\cite{chen2021evaluating}, HumanEval-XL~\cite{peng2024humaneval}, and CoNaLa~\cite{yin2018learning} provide useful test cases for assessing the ability of \toolname in evaluating code correctness, they fall short when assessing more nuanced aspects like helpfulness, readability, and alignment with human judgment. Consequently, a critical limitation in existing literature on \toolname of SE is their reliance on small-scale datasets to measure human alignment~\cite{wang2025can, wu2024can}, often comprising only a few hundred samples. For example, Wang et al.~\cite{wang2025can} conducted experiments on three SE tasks, i.e., Code Translation, Code Generation, and Code Summarization, using a total of 450 samples. Similarly, Ahmed et al.~\cite{ahmed2024can} evaluated code summarization using 420 samples. Such limited datasets threaten the generalizability of findings and contribute to inconsistent empirical results across studies. CodeVisionary~\cite{wang2025codevisionary} conducted experiments on 363 human-annotated code generation samples. This scarcity is not merely an issue of effort; it points to a more fundamental difficulty in defining a ``gold standard'' for these subjective tasks.

\noindent \textbf{Limitation 2: Neglect of Indeterminacy, Uncertainty, and Preferences.}
However, a high-quality benchmark for evaluating \toolname in SE is indeed challenging to develop. A significant limitation in current SE benchmarks of LLM-as-a-Judge systems is the neglect of three intertwined concepts: \emph{Rating Indeterminacy}, \emph{Evaluator Uncertainty}, and \emph{Evaluator Preferences}. Current validation practices often operate on the flawed assumption that a single, objective ground truth exists for subjective software engineering tasks. This oversimplification ignores that genuine expert evaluation is not a monolithic process, but is instead shaped by three critical factors: \emph{Rating Indeterminacy}, \emph{Evaluator Uncertainty}, and \emph{Evaluator Preferences}.

\begin{itemize}[leftmargin=*]
    \item \emph{Rating Indeterminacy}: A task is indeterminate~\cite{pavlick2019inherent} when it lacks a single, objectively correct rating. Indeterminacy allows for multiple reasonable assessments, even from one SE expert. This is not a flaw in the evaluation process but an inherent characteristic of many evaluation qualities in software engineering tasks. This property arises from two primary sources: vagueness and ambiguity. Vagueness stems from imprecise definition of a criteria. For example, concepts like ``a good design,'' ``maintainability,'' or ``readability'', which lacking a universally accepted definition. Ambiguity arises from a lack of sufficient context. However, current SE literature of \toolname often obscure this reality by aggregating human ratings into a single ``gold label,'' thereby erasing legitimate alternative perspectives. The issue is exacerbated by \textbf{forced-choice elicitation}, which requires evaluators to select only one option. As a result, this practice is likely to biased evaluations when humans and LLMs resolve indeterminacy differently.

    \item \emph{Evaluator Uncertainty}: Evaluator uncertainty refers to the inherent variability in human judgments, where different experts may disagree on the same artifact. An indeterminate task often forces the judge to be uncertain.

    \item \emph{Evaluator Preferences}: The issue of indeterminacy is further complicated by evaluator preferences. Preferences~\cite{liu2024learning} are the specific, often unwritten, conventions of a team or individual regarding coding style, architectural patterns, documentation standards, etc. Indeterminacy creates the space for these preferences to become the deciding factor.
\end{itemize}

\noindent \textbf{Limitation 3: Inadequacy of Traditional Alignment Metrics.} Current SE literature of \toolname often rely on metrics like inter-rater agreement (IRA) coefficients (e.g., Krippendorff's $\alpha$~\cite{krippendorff2011computing}, Spearman's $\rho$~\cite{wissler1905spearman}) to measure the human-LLM alignment. However, these metrics are possibly ill-suited for comparing LLMs' judgments to human evaluations because they were originally designed to measure agreement between human raters. The reasons are as follows:

\begin{itemize}[leftmargin=*]
    \item These metrics operate on the flawed premise that a single correct answer exists for subjective tasks like rating code readability. They penalize the natural variation in human perception by forcing researchers to aggregate diverse judgments into one ``gold standard'' label (e.g., a majority vote), discarding valuable information about the distribution of expert opinion.
    \item Metrics like Krippendorff's $\alpha$~\cite{krippendorff2011computing} were originally designed to measure the reliability of human labeling and it assumes all raters are interchangeable peers. This core assumption is clearly violated when comparing a LLM to human experts, as humans provide the ground truth labels.
    \item IRA metrics are designed to account for random, non-reproducible errors typical of human raters~\cite{krippendorff2011computing,elangovan2024beyond, wissler1905spearman}. However, the errors made by LLMs are consistent and reproducible, even if hard to predict. Traditional metrics are ill-equipped to detect or correctly analyze this kind of systematic divergence between a human and LLMs.
    \item These IRA metrics do not provide a clear threshold for what constitutes a good performance. A Spearman's correlation of 0.5 might be considered moderate, but it doesn't tell a researcher if an LLM is a valid replacement for a human evaluator or what specific gaps in judgment exist.
\end{itemize}

\noindent \textbf{Limitation 4: Inconsistent Empirical Findings. }
Consequently, the field is currently hampered by inconsistent and contradictory empirical findings~\cite{wang2025can, wu2024can}, which is a direct symptom of the previously discussed limitations. A few works reached varying conclusions due to discrepancies in dataset selection, prompting strategies, and LLM configurations.
For instance, Wang et al.~\cite{wang2025can} found that traditional evaluation metrics (e.g., ROUGE~\cite{lin2004rouge}, METEOR~\cite{banerjee2005meteor}, and ChrF++~\cite{popovic2017chrf++}) significantly outperformed LLM-as-a-Judge methods when aligning with human judgment for code summarization. In contrast, Wu et al.~\cite{wu2024can} reported that the \toolname method surpassed conventional metrics in code summarization. These conflicting results highlight the need for a standardized, large-scale empirical study to enable fair and meaningful comparisons across studies.

\noindent \textbf{Limitation 5: Limited Empirical Findings on Bias of \toolname systems in SE.} The evaluation of software artifacts by LLMs is susceptible to biases~\cite{9825855,9609175}. LLMs are known to exhibit various cognitive biases, including Position Bias, where responses are favored based on their order~\cite{li-etal-2024-split}; Verbosity Bias, a preference for longer answers regardless of quality~\cite{jiao2024enhancing}; and Egocentric Bias, the tendency to overrate AI-generated solutions~\cite{ye2024justice}.
The SE community has started to investigate these cognitive biases. For instance, Crupi et al.~\cite{crupi2025effectiveness} analyzed ``self-bias,'' where an LLM judge prefers its own generated content. However, existing work on LLM-as-a-Judge biases in SE is still in its early stages and overlooks more complex biases, such as Auxiliary Information Induced Biases~\cite{li2025curse}. These biases arise when an LLM judge is provided with external context, like reference answers or scoring rubrics. While intended to improve accuracy, this information can paradoxically make the judge vulnerable to new forms of bias and manipulation.

\vspace{0.1cm}
\noindent\textbf{\textit{Opportunity: Develop Comprehensive and Diverse Benchmarks.}} Future research should work on creating large-scale, multi-dimensional benchmarks that capture the complexity of real-world software engineering tasks. Key steps include:
\begin{enumerate}[leftmargin=*]
    \item \textit{Expert Annotations and Quality Control:} Engage a diverse pool of expert programmers to provide high-quality annotations. This annotation process should focus more on the nuanced criteria beyond code correctness, such as readability, helpfulness, and maintainability. In the beginning, define clear annotation guidelines that describe these nuanced dimensions to eliminate ambiguity.
    During the process, cross-validation and consensus mechanisms should be conducted to ensure the reliability of the annotations and improve the clarity of the guidelines.

    \item \textit{Dataset Expansion:} Expand existing datasets to include a larger volume of high-quality expert-annotated samples. These benchmarks should encompass SE tasks at varying difficulty levels, various programming topics, and multiple programming languages. Additionally, new benchmarks should be extended to evaluate a broader spectrum of SE artifacts. For instance, expert evaluations could be gathered for software requirements documentation, system design, and API documentation.
    \item \textit{Updating Human Preferences:} In the long term, the annotation process should be enriched to include explicit human preferences to reflect evolving SE practices.
\end{enumerate}

\noindent\textbf{\textit{Opportunity: Establish a Standard and Nuanced Evaluation Protocol.}} To overcome the current challenges, the community must move beyond simplistic aggregate scores and adopt a more rigorous evaluation paradigm. This requires establishing a standard protocol designed to handle the inherent subjectivity of software engineering tasks and provide a clearer, more comparable picture of an LLM's performance. Such a protocol should be built on the following principles:

 \begin{itemize}[leftmargin=*]
    \item \textit{Acknowledge the Subjectivity and Uncertainty:} Instead of aggregating diverse human ratings into a single ``gold label,'' the protocol should mandate that benchmarks capture the full distribution of expert opinions. This paradigm shift treats the range of human judgments not as noise, but as the ground truth itself. This approach naturally accommodates tasks with inherent indeterminacy (e.g., rating ``readability'') and makes evaluator preferences an explicit part of the evaluation landscape rather than an unacknowledged variable.
    \item \textit{Adopt Distribution-Aware Alignment Metrics:} The protocol must advocate for metrics that compare the distribution of LLM judgments against the distribution of human judgments. This moves away from metrics like Krippendorff's $\alpha$ or Spearman's $\rho$, which fundamentally rely on a single correct answer. Instead, research should focus on adopting or developing metrics (e.g., ~\cite{menendez1997jensen}, Earth Mover's Distance~\cite{rubner2000earth}) that quantify how well an LLM's range of assessments mirrors the range of human expert assessments. This correctly reframes the goal from matching a single label to replicating the spectrum of expert reasoning.
    \item The protocol should require researchers to analyze and report performance separately on data subsets with high and low human agreement. By doing so, the community can distinguish an LLM's ability to handle clear, unambiguous cases from its performance on more contentious, indeterminate ones. This practice will expose the performance gaps that single aggregate scores often hide and help explain why different studies, using datasets with varying levels of uncertainty, have reached contradictory conclusions.
 \end{itemize}

\vspace{0.1cm}
\noindent\textbf{\textit{Opportunity: Conduct Comprehensive Empirical Evaluations in SE.}} Ultimately, the benchmarks and new evaluation protocol will lay a solid foundation for understanding the strengths and weaknesses of LLM-as-a-Judge systems in SE. The community will be positioned to conduct a large-scale, comprehensive empirical evaluation. The goal is to create a reliable, evidence-based foundation that provides clear and actionable guidance for both researchers and practitioners. This comprehensive evaluation should consider the following three core pillars:
\begin{enumerate}[leftmargin=*]
    \item \textit{Systematic Comparison Across Diverse SE Tasks.} The study must systematically assess the performance of \toolname across a wide array of software engineering tasks. This would go beyond conventional tasks like code generation to include evaluations of software design documents, API usability, and system requirements. This broad scope will help identify where LLM judges excel and where they still fall short.

    \item \textit{Rigorous Analysis of Design Variables.} Studies should rigorously investigate how different design choices impact an LLM's judgment. Key variables to analyze include:
    \begin{itemize}
        \item \textit{Prompting Strategies:} Evaluating the effectiveness of different prompting techniques, such as Chain-of-Thought, reference-based rubrics, and few-shot examples.
        \item \textit{LLM Configurations:} Comparing various models (e.g., GPT-4, Claude 3, Llama 3), model sizes, and parameters like temperature to understand their effect on evaluation quality and consistency.
    \end{itemize}

    \item \textit{More Comprehensive Studies on Biases in \toolname in SE.} Future research should conduct comprehensive empirical studies to better understand more complex biases in \toolname systems for SE artifacts. For example, the auxiliary information-induced biases, where external context paradoxically leads to flawed evaluations. For example, studies can systematically investigate reference-induced biases, such as ``solution fixation,'' where a judge penalizes valid solutions that deviate from a given reference. Furthermore, research should explore whether the severity of these biases correlates with task complexity and whether advanced reasoning models are more susceptible than general-purpose models, particularly when their judgments rely on flawed auxiliary information.

    \item \textit{Creation of Actionable Guidance and Baselines.} The ultimate output should not just be a collection of results but a set of clear, actionable guidelines for the community. This would include recommendations on which models and prompting strategies are best suited for specific SE tasks, as well as establishing strong, reproducible baselines. This will enable future research to build on a common foundation, ensuring that new findings are comparable and contribute to a cumulative body of knowledge.
\end{enumerate}

\subsection{Better Judgment From Internal Intelligence}

\noindent \textbf{Limitation 6: Inadequate SE Domain-Specific Expertise in LLMs.} While ideally, human evaluations should be conducted by experienced developers. Like human evaluators, LLM must exhibit a deep understanding of the task and the relevant domain knowledge to provide an accurate evaluation.
Recent research highlights that LLMs have remarkable coding abilities, which builds confidence in their ability to assess tasks like code summarization and generation.
However, research also indicates that some complex coding tasks are still challenging for LLMs~\cite{zhuo2024bigcodebench, zhao2024commit0}. For instance, Zhao et al.~\cite{zhao2024commit0} observed that current LLMs cannot generate a complete library from scratch, which also implies LLMs' ability to evaluate library quality is still limited. Similar limitations extend to other critical areas in software engineering, such as software design, formal verification, distributed systems debugging, etc.
Beyond gaps in SE-specific knowledge, LLMs may also struggle with discerning subtle differences among alternative solutions. 
As Zhao et al.~\cite{zhao-etal-2025-codejudge} highlighted, an LLM's ability to generate correct code does not guarantee that it can effectively evaluate alternative implementations for the same task.

\noindent \textbf{Limitation 7: The Mono-modal Focus of Current LLM-as-a-Judge Evaluation.} Current LLM-as-a-Judge research in SE is overwhelmingly mono-modal. It is focusing almost exclusively on artifacts that exist as text and source code. This creates a significant blind spot, as software engineering is an inherently multi-modal discipline. Developers constantly work with visual artifacts like UML diagrams, architectural sketches, GUI mockups, and screenshots in bug reports. Current LLM judges cannot ``see'' these artifacts, they must rely on ``flattened'' textual descriptions, a process which can lose critical and structural information that is immediately apparent to a human expert.

\vspace{0.1cm}
\noindent\textbf{\textit{Opportunity: Enhance the SE Expertise of LLMs.}} Future research can strengthen the SE domain-specific expertise of LLMs. A key approach is training LLMs on richer and more high-quality SE datasets~\cite{liu2024datasets}. Enhancing SE expertise in LLMs offers several benefits. For instance, LLMs with stronger formal verification capabilities can more accurately assess code correctness. Further, research can expand \toolname's applicability to a broader range of SE artifacts, such as Docker files, UML diagrams, and formal specifications. 

\vspace{0.1cm}
\noindent\textbf{\textit{Opportunity: Embedding Expert Tacit Knowledge.}}
Another critical research direction for enhancing judgment involves capturing the tacit and procedural knowledge that human experts develop over years of experience, i.e., their intuition and unspoken decision-making processes. We can employ methods such as structured interviews~\cite{segal2006structured}, think-aloud protocols~\cite{zhang2019think}, and cognitive task analyses~\cite{schraagen2000cognitive}, where experts are asked to articulate their reasoning while conducting real-world evaluations. This approach enables the systematic extraction of procedural knowledge that is typically not documented in software artifacts. Once captured, this knowledge can be integrated into LLM training through techniques like reinforcement learning~\cite{bai2022training} or neuro-symbolic methods~\cite{kwon2023neuro}. 

\vspace{0.1cm}
\noindent\textbf{\textit{Opportunity: Incorporation of Multi-modal Assessment.}} Multi-modal LLMs (MLLMs)~\cite{qin2024multilingual, wu2024visionllm} are capable of reasoning over images, text, and code simultaneously, which presents a significant opportunity for the LLM-as-a-Judge paradigm~\cite{chen2024mllm}. The development of MLLM judges can enable a holistic evaluations, that is more closely mimicking the multi-modal workflow of a expert developer. MLLM judges can enable a set of cross-modal verification that are currently beyond the capabilities of text-only models. For example:
\begin{itemize}[leftmargin=*]
    \item \textit{Requirements-to-Design Conformance:} Verifying that a system's design, as captured in a UML diagram or GUI mockup, faithfully implements the specified textual requirements.
    \item \textit{Architectural Drift Detection:} Analyzing a high-level architectural diagram and comparing it against the code's dependency graph to identify deviations from the intended architecture.
    \item \textit{GUI Implementation and Usability Testing:} Evaluating a user interface by comparing application screenshots against design mockups to verify the precise implementation of visual elements and assess adherence to established UI/UX principles.
\end{itemize}

\subsection{Augmenting Judgment with External Insights}

\noindent \textbf{Limitation 8: Limited Integration of External SE Tools.} While early LLM-as-a-Judge methods depended exclusively on the model's internal knowledge , this approach is insufficient for comprehensive evaluation, as it lacks access to crucial runtime, linting, and visual information. Pioneering research has begun to address this by developing agent-based frameworks~\cite{wang2025codevisionary,zhuge2024agent}, like CodeVisionary~\cite{wang2025codevisionary}, which can interact with and leverage external tools such as static linters, execution environments, and unit testers. However, this practice is still in its early stages and not yet widespread. The full potential of a synergistic relationship between the LLM judge and the rich ecosystem of specialized software engineering tools remains largely untapped.

\vspace{0.1cm}
\noindent\textbf{\textit{Opportunity: Integrate Diverse SE Tools for Richer Evaluations.}} Building on the initial success of tool-augmented agents, a significant opportunity lies in creating frameworks where the LLM judge acts as an orchestrator, intelligently leveraging a wider array of external SE tools to form its judgments. While pioneering work like CodeVisionary has already integrated foundational tools like static linters and execution environments, future research can expand this integration to be far more comprehensive and sophisticated. For instance, the evaluation pipeline could incorporate formal verification frameworks~\cite{shi2025synthesizing}, model checkers~\cite{david2007model}, and performance profilers to assess code efficiency, including execution time and memory usage~\cite{ShiYL25,yang2024acecode}.

\vspace{0.1cm}
\noindent\textbf{\textit{Opportunity: Intelligent Human-in-the-Loop Collaboration.}}
For tasks that cannot be reliably automated by LLMs, we need to develop collaborative and 
interactive methods to include human oversight.A key component of such a system is enhancing LLMs with the ability to assess their own confidence. When a judge generates a low-confidence rating for a nuanced or high-stakes task, it should automatically flag the case for human review. This creates a structured workflow where LLMs handle high-volume, routine checks (e.g., style, simple correctness), allowing human experts to focus their limited time on the most complex evaluations. Also, identifying the optimal ratio of LLM 
evaluators to human experts remains another challenge. Furthermore, this collaboration should not be a one-way street. The judgments and corrections provided by human experts must be collected as high-quality data to continuously fine-tune the LLM judge. This establishes a virtuous feedback loop, progressively aligning the AI's reasoning with that of human experts over time and addressing the challenge of finding the optimal balance between automated and manual evaluation.

\subsection{Building Robust Judges}

\noindent \textbf{Limitation 9: Insufficient Research on Adversarial Threats and Defensive Methods in SE.}
Adversarial attacks on LLM-as-a-Judge systems pose a significant threat to software integrity. Current research in this area is still nascent. While frameworks like RobustJudge~\cite{li2025llms} have initiated the study of adversarial attacks in SE, their focus has largely been on overt methods such as prompt injection~\cite{li2025llms}. These attacks, often considered ``low-hanging fruit,'' are typically detectable and can be mitigated with straightforward defenses like input sanitization. However, this emphasis on overt attacks overlooks the more dangerous class of sophisticated, semantics-preserving manipulations. These attacks subtly alter software artifacts to deceive the judge into accepting flawed or malicious code~\cite{yang2024robustness,yang2022natural}. For instance, an attacker might introduce convoluted but functionally correct logic that harms maintainability, or using intentionally confusing variable names to mislead the judge. Because such subtle manipulations mimic legitimate coding practices, they are far more difficult to detect automatically. Consequently, this leaves LLM judges vulnerable to advanced threats that have serious implications for code quality and security.

\vspace{0.1cm}
\noindent\textbf{\textit{Opportunity: Uncovering Vulnerabilities via Advanced Adversarial Testing.}}
The nascent state of security research presents a critical opportunity to move beyond testing for overt attacks and systematically evaluate the robustness of LLM judges against more sophisticated threats. Future work should focus on developing and applying advanced adversarial testing methods to uncover hidden vulnerabilities. This includes creating optimization-based attacks, which use automated refinement to craft subtle inputs that probe for weaknesses in a model's reasoning. Another critical area for adversarial testing is evaluating resilience against persistent backdoor attacks~\cite{10431665,gong2024baffle}, where a judge's training data is poisoned with innocuous triggers. For example, testing could involve inserting a specific comment style as a trigger to see if it activates a biased judgment. Such testing is crucial because, unlike overt attacks, these vulnerabilities are stealthy and bypass standard defenses, posing a persistent threat. A rigorous adversarial testing framework is essential for understanding these complex failure modes before effective defenses can be developed.

\vspace{0.1cm}
\noindent\textbf{\textit{Opportunity: Fortifying Judges Against Advanced Threats.}}
Future research should focus on building robust LLM judges capable of withstanding sophisticated adversarial threats. This requires a two-pronged strategy that combines proactive defense mechanisms with efforts to enhance the model's intrinsic security. On the one hand, researchers must develop defenses against advanced attacks like optimization-based methods and persistent backdoors. This involves creating dynamic defense mechanisms, such as anomaly detection~\cite{song-etal-2025-confront,10431665}, and new techniques for data sanitization and model analysis to neutralize stealthy vulnerabilities introduced through data poisoning. On the other hand, a more powerful approach is to enhance the judge's intrinsic robustness, reducing reliance on external defenses. A promising direction is adversarial fine-tuning, where models are explicitly trained on datasets of successful attacks~\cite{gong2022curiosity,9825775,xhonneux2024efficientadversarialtrainingllms} or on pairs of adversarial and ``purified'' code to learn to identify and resist manipulation patterns. Furthermore, advanced prompting techniques can foster self-awareness and skepticism in the judge. For instance, incorporating a Chain-of-Thought for security analysis or implementing self-correction cycles would enable the model to critique its own judgments for potential adversarial influence before delivering a final evaluation. By integrating both external defenses and intrinsic hardening, the community can develop more resilient and trustworthy LLM-as-a-Judge systems.

\section{A Vision for the Future}

This section presents our vision for LLM-as-a-Judge, illustrating the future capabilities outlined in our roadmap for evaluating LLM-generated software artifacts.

\subsection{Calibrated Expert Judgment}

Beyond assessing functional correctness, the LLM Judge will deliver nuanced verdicts that emulate the assessments of an expert human developer. It will serve as a cost-effective surrogate for human experts, capable of evaluating complex qualities such as code maintainability and the presence of code smells. A significant leap will be in the domain of security; while current models often lack the precision for reliable vulnerability detection, the future LLM Judge will provide accurate judgments in this critical area. This expert-level feedback can be provided in real-time, enabling the rapid and iterative refinement of generated artifacts at scale.

\subsection{Preference-Aware, Multi-Perspective Feedback}

The visionary LLM Judge will treat subjectivity as a configurable feature. By adopting distinct evaluation personas, it will emulate the preferences of different domain experts and assess an artifact from multiple valid viewpoints. This approach directly addresses the challenges of Rating Indeterminacy and Evaluator Preferences. Such targeted feedback allows evaluation to be tailored to specific project needs, overcoming the limitations of a single, generic ``gold standard''.

\subsection{Cross-Modal Holistic Assessment}

Software engineering is an inherently multi-modal discipline. By 2030, the LLM Judge will be a Multi-modal LLM, capable of reasoning across diverse artifact types to ensure their coherence. For example, an MLLM Judge could simultaneously analyze a generated UML class diagram and its corresponding code implementation to identify discrepancies between the design and its implementation. This cross-modal verification, impossible for current text-only models, represents a major leap toward achieving holistic and reliable evaluation.

Ultimately, this visionary judge will provide a comprehensive, multi-faceted, and adaptable evaluation. It will ensure the quality, security, performance, and internal consistency of LLM-generated software, far exceeding the capabilities of today's automated metrics.

\section{Conclusion}

This forward-looking SE 2030 paper presents an overview of the current landscape of LLM-as-a-Judge systems in software engineering. We conduct a systematic review of 42 primary studies in the field, and map the applications across the software development lifecycle. Further, we identify the limitations of \toolname in SE while highlighting the associated opportunities and presenting a future vision for its adoption in the community. To realize this vision by 2030 and beyond, we propose a research roadmap outlining the specific paths and potential solutions that the research community can pursue. We propose a list of short-term goals and long-term goals. Through these efforts, we aim to encourage broader participation in the LLM-as-a-Judge research journey.

\bibliographystyle{ACM-Reference-Format}
\bibliography{reference}

\end{document}